\documentclass[twocolumn,prb,floatfix]{revtex4}
\usepackage{amsmath}
\usepackage{amsfonts}
\usepackage{amssymb}
\usepackage{bm}
\usepackage{color}
\usepackage{graphicx}

\begin{document}

\title{Coarse graining of master equations with fast and slow states}

\author{Simone Pigolotti$^1$ and
Angelo Vulpiani$^2$} 

\affiliation{$^1$The Niels Bohr Institute, The Niels Bohr
  International Academy, Blegdamsvej 17, DK-2100 Copenhagen, Denmark\\
  $^2$ Dipartimento di Fisica and INFN Universit\`a di Roma ``La
  Sapienza'', Piazzale A.\ Moro 2, I-00185 Roma, Italy}

%\date{}
\begin{abstract}
We propose a general method for simplifying master equations by
eliminating from the description rapidly evolving states. The physical
recipe we impose is the suppression of these states and a
renormalization of the rates of all the surviving states. In some
cases, this decimation procedure can be analytically carried out and
is consistent with other analytical approaches, such as in the problem
of the random walk in a double-well potential. We discuss the
application of our method to nontrivial examples: diffusion in a
lattice with defects and a model of an enzymatic reaction outside the
steady state regime.
\end{abstract}

\maketitle

%%%%%%%%%%%%%%%%%%%%%%%%%%%%%%%%%%%%%%%%%%%%%%%%%%%%%%%%%%%%%%%%%%%%%%
\section{Introduction}
\label{intro}
%%%%%%%%%%%%%%%%%%%%%%%%%%%%%%%%%%%%%%%%%%%%%%%%%%%%%%%%%%%%%%%%%%%%%%

Many problems of relevance in physics, chemistry and biology are
appropriately described as systems of interacting, discrete entities
which evolve according to stochastic rules.  These entities may be not
all equal, for example they may be molecules of different chemical
species. When a) there is spatial homogeneity and other continuous
degrees of freedom are irrelevant for the description, and b) the
Markovian property holds, such systems are well described by master
equations,
\begin{equation}
\frac{d}{dt}P_n= \sum_m\left( W_{mn}P_m-W_{nm}P_n\right).
\end{equation}
Here $n$ denotes the state of the system, $P_n(t)$ is the probability
of being in state $n$ at time $t$, and $W_{mn}$ is the transition rate
from state $m$ to state $n$.  For instance in multi-species chemical
systems, $n$ is a vector $(n_1,\ldots ,n_s)$, the component $n_j$ being
the number of molecules of the $j$-th specie.

The description of chemical and biological processes in terms of
master equations is usually rather accurate. However, it can have
rather severe problems, even in the numerical treatment, if the number
of states is large and overall if many degrees of freedom are
involved.  When the number of particles is large enough, a possibility
is to describe the dynamics in terms of concentrations: there are
several methods to move from a master equation to a partial
differential equation, like the Van Kampen system size expansion
\cite{vankampen}.  The most common result of such a procedure is a
Fokker-Plank equation.

On the other hand there are systems which are fundamentally discrete
and, therefore, the continuous approximation may give answers which are
quite different from that of the discrete case, like enzymatic
reactions which are often studied in the limit where one has many
substrate molecules but very few enzymes \cite{stefanini}. Other
notable examples are genetic systems working with low molecules copy
numbers \cite{elowitz, kardar} and ecological systems close to the
extinct absorbing state \cite{mckane}.

A common features of many of these problems which can be used to
simplify the description is the presence of many relevant
timescales. Sometimes the timescale of the interactions is fast, but
due to the many degrees of freedom the timescale of the global
dynamics is much slower. This is the case of protein folding: while
the elementary time scale of vibration of covalent bonds is $\sim
10^{-15} s$, the folding time for a protein may be of order of
seconds.  Another problematic situation is when the degrees of freedom
are not so many, but the interactions are of diverse nature, resulting
in entries in the transition matrix $W_{mn}$ of very different orders
of magnitude. An example of the latter case comes from many cell
regulatory systems: proteins in a cell can interact chemically, again
on molecular timescales, and via transcription regulation, on
timescales of seconds or even minutes.  In all these situations one
says that the system has a \textit{multiscale} character
\cite{eenquist}.

The necessity of treating the ``slow dynamics'' in terms of effective
equations is both practical (even modern supercomputers are not able
to simulate all the relevant scales involved in certain difficult
problems) and conceptual: effective equations are able to catch some
general features and to evidence key controls and basic ingredients
which can remain hidden in the detailed description.  The study of
multiscale problems has a long history in science: perhaps the first
example is the study, due to Newton, of the precession of the
equinoxes, which was basically a special version of the averaging
method in mechanics \cite{arnold}.  In fluid dynamics an example of
the multiscale procedure is the derivation of an effective Fick
equation for the large scale and long time behavior starting from the
transport equation \cite{biferale}: the diffusion tensor depends,
often in a non intuitive way, on the velocity field.  Finally, in
quantum mechanics, since the nuclei are much heavier than the
electrons, one can simplify the treatment ``splitting'' the electronic
and nuclear degrees of freedom, like in the Born-Oppenheimer
approximation \cite{ziman}, or in its modern generalization due to Car
and Parrinello \cite{car}.

Multiscale systems described by master equations can be difficult to
deal with.  Gillespie's stochastic simulation algorithm
\cite{gillespie}, which is an exact and well established method for
numerical simulations of master equations, is not very efficient to
simulate multiscale systems. The reason is that Gillespie algorithm
treats fast and slow dynamics on equal footing, while if one is
interested in the slow dynamics it is usually not necessary (and
computationally demanding) to exactly integrate the fast dynamics. In
recent years, several authors introduced modifications and
approximations of the Gillespie algorithm that allow to improve its
efficiency when applied to multiscale problems. Many approaches have
been developed like the quasi-steady state approximation
\cite{raoarkin,goutsias}, fast variables elimination methods
\cite{frankowicz,pineda}, finite state projection techniques
\cite{munsky,peles}, or continuous approximations of fast variables
\cite{haseltine,cao}.

The method developed in this paper follows a different idea. We want
to study master equation containing states with fast and slow dynamics
(evolving with different characteristic times). Our approach is to
write down an effective master equation describing the evolution of
the slow states only.  The physical recipe we are going to impose is
that, looking at the system on a slow timescale, the time spent on the
fast states can be neglected. These states are thus eliminated from
the description and this brings to new effective transitions among the
slow states.  Clearly this approximation gives better results when the
separation of timescales between fast and slow states becomes
large. Our procedure, by reducing the number of states, may allow for
a large gain in simulation time. However, our main goal is to obtain
in a systematic way a simple description of a system obeying a master
equation on a slow timescale.  Indeed, we will show that in some cases
our method does not correspond only to a numerical recipe, but allows
for analytical predictions.

The plan of the paper is the following.  In Section
\ref{methodsection} we introduce our method.  Section
\ref{onedsection} is devoted to the one dimensional examples: a random
walk in a double-well potential and a random walk in a lattice with
defects. Section \ref{enzymesection} we apply our method to a common
model of enzymatic reaction and show how it can predict its behavior
far from the steady state regime. Conclusions and perspectives are in
section \ref{conclsection}.

\section{The method}\label{methodsection}

\begin{figure}[h]
\begin{center}
\includegraphics[width=8.3cm]{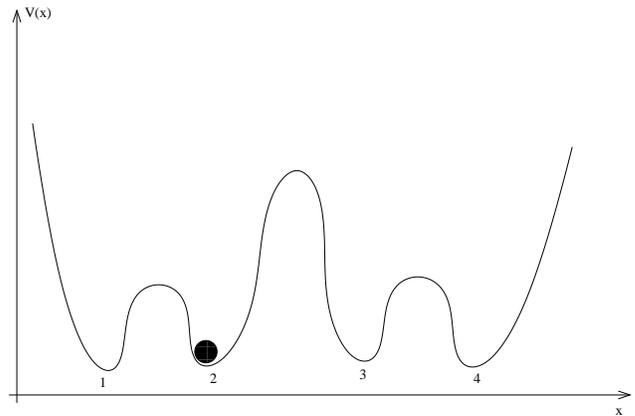}
\end{center}
\caption{Schematic example: a particle confined in a four-well
potential.}\label{figesempio}
\end{figure}

Let us introduce the intuitive idea of the method with a schematic
example.  We consider the motion of a Brownian particle in a potential
$V(x)$ having $4$ minima, like the one sketched in
Fig. \ref{figesempio}.  

When the temperature $T$ is not too large the probability distribution
function of the particle is concentrated only around the the minima
$x_1, x_2, x_3$ and $x_4$.  Therefore, instead of the complete
Fokker-Planck description, it is sensible to study the system in terms
of a Master equation with $4$ states: we say that the system is in the
state $i=1, 2, 3, 4$ when $x(t)$ is close to $x_i$.  Moreover, we
consider the case in which the barriers between states $1$, $2$ and
$3$, $4$ have the same height $\Delta V_1$, being smaller than the
height $\Delta V_2$ of the barrier between states $2$ and $3$.  In
this case, one has two very different typical times: the transition
time between states $1$, $2$ and $3$, $4$ $\tau_1 \sim e^{ \Delta
V_1/T}$ and the transition time between $2$ and $3$, $\tau_2 \sim e^{
\Delta V_2/T}$.  If one is interested to properties on times much
longer than $\tau_1$, it is quite natural to devise a model of the
system with only two states, say $A$ which includes $1$ and $2$ and
$B$ which includes $3$ and $4$. This simple example suggests how the
study of a system on a slower timescale may lead to a reduction in the
number of states one has to consider.

Consider now a generic master equation
\begin{equation}\label{mastereq}
\frac{d}{dt}P_n= \sum_m\left( W_{mn}P_m-W_{nm}P_n \right).
\end{equation}

The standard way to simulate the above equation would be of course the
Gillespie algorithm \cite{gillespie}, which is exact and does not
imply a choice of a timestep. On the other hand, we would like to
select a (slow) timescale, so let us discuss what happens if we
integrate naively the equation, for example discretizing the time via
the Euler algorithm:

\begin{equation}\label{markov}
P_n^{t+\Delta t}=P_n^{t}+\Delta t \sum_m\left( W_{mn}P_m-W_{nm}P_n\right) = \sum_m
A_{nm}^{\Delta t} P_m
\end{equation}

In this case, the application of the Euler algorithm corresponds to
approximate the master equation with a Markov chain, defined by the
Markov transition matrix $\hat{A}^{\Delta t}$
\begin{equation}\label{markovmatrix}
A_{mn}^{\Delta t}=\left\{\begin{array}{cc}\Delta t\ W_{nm}\quad & m\neq n\\
1-\Delta t W_n^{out} \quad & m=n\end{array} \right.
\end{equation}
where we introduced for convenience of notation the total out-rate of
state $n$,
\begin{equation}
W_n^{out}=\sum_{k\neq n} W_{nk}.
\end{equation}
When integrating Eq.(\ref{mastereq}), $\Delta t$ should be chosen to
be small compared to the timescales of the system dynamics.  However,
notice that Eq.(\ref{mastereq}) and (\ref{markov}) share the same
stationary condition independently on $\Delta t$. This means that they
have the same stationary state for any $\Delta t$, even though they
can have in principle quite different dynamics when $\Delta t$
increases. A natural question is what happens when $\Delta t$ becomes
large. The problem one encounters is that the diagonal element of the
matrix (\ref{markovmatrix}) can become negative. This starts happening
when
\begin{equation}\label{dtstates}
\Delta t > \max\limits_n\left\{ (W_n^{out})^{-1} \right\}.
\end{equation}
On the other hand, when $\Delta t \ll \max_n\{ (W^{out}_n)^{-1}\}$, the
Markov chain and the original master equation are closely related. Not
only they share the same stationary state, but one can easily write a
relation between their eigenvalues, which are in a one to one
correspondence. We can write the Markov matrix $\hat{A}^{(\Delta t)}$:
\begin{equation}
\hat{A}^{(\Delta t)}=\hat{I}+\Delta t \hat{W}^\dagger
\end{equation}
where $\hat{I}$ is the identity and $\hat{W}$ is the matrix having as
non-diagonals elements the transition rates and on the diagonal minus
the total out-rates. This implies that, calling $\lambda_A$ and
$\lambda_W$ the eigenvalues of the matrices $\hat{A}$ and $\hat{W}$,
one has
\begin{equation}\label{eigenrelation}
\lambda_A^{(\Delta t)}=1+\Delta t \lambda_W.
\end{equation}
Another way of seeing it is that $\hat{A}^{(\Delta t)}$ is the first-order
expansion of the evolution operator of the master equation
$\exp(\hat{W}^\dagger\Delta t)$. This implies the error one makes on the
eigenvalues is order $O(\lambda_w^2\Delta t^2)$ and should be
negligible for the slower modes when $\Delta t$ is not too large.

\subsection{Decimation procedure}

\begin{figure}[h]
\begin{center}
\includegraphics[width=8.3cm]{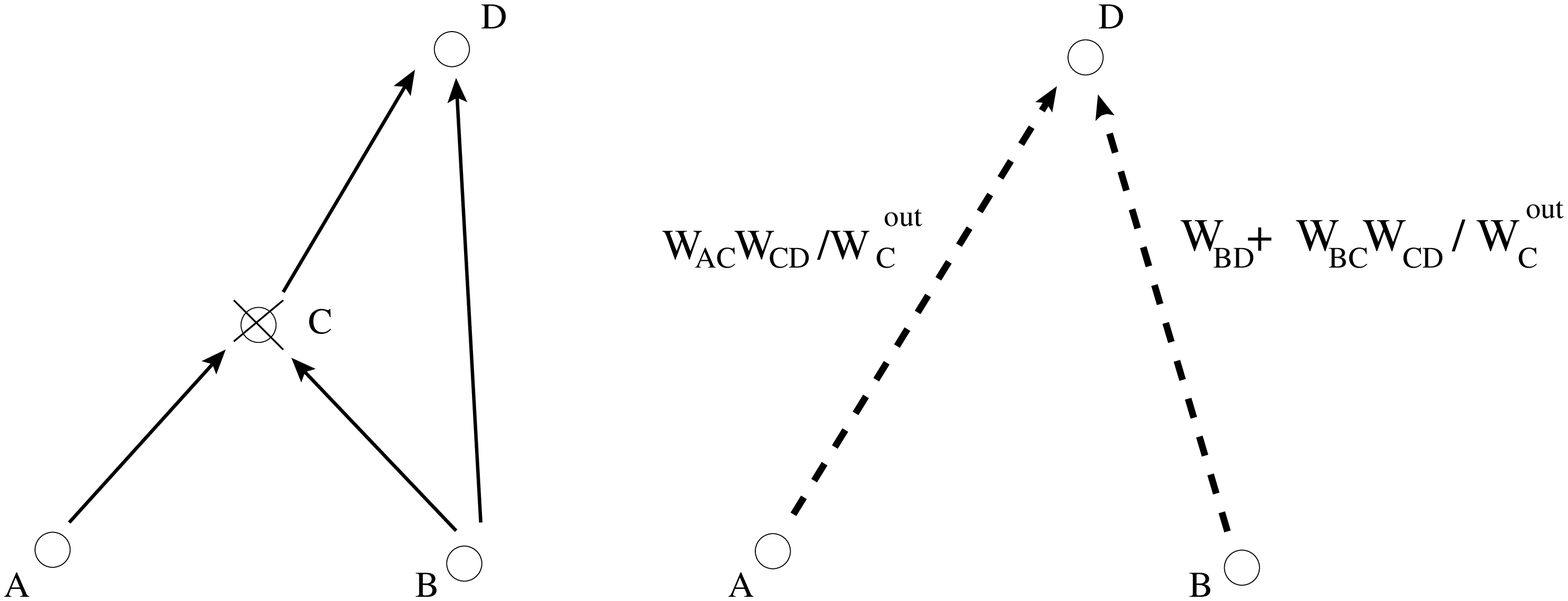}
\end{center}
\caption{Schematic example of the transition rates before and after
the decimation in a master equation with $4$ states. Here, the state
$C$ on the left is a fast states and is eliminated from the
description. On the right we represent the renormalized transition in
terms of the original rates. The new transition rate from $B$ to $D$
contains the original one plus an additional contribution coming from
the elimination of $C$. The transition rate from $A$ to $D$, equal to
zero in the original graph, contains only the effect of the
decimation.}\label{figscheme}
\end{figure}

Eq. (\ref{dtstates}) suggests a strategy to identify ``fast states''
on a given timescale. The idea is that, every time a generic state $n$
is reached, the average time spent in it is $(W_n^{out})^{-1}$. As a
consequence, if we choose the parameter $\Delta t$ representing the
smaller timescale we aim to describe, then all states $n$ having
$W_n^{out}>\Delta t^{-1}$ should not enter into the description.  In
order to eliminate them, the physical recipe we impose is simply
that the time spent on these states is zero. In this way, the states
disappear from the dynamics and the transition rates from a generic
state $k$ to a generic state $j$ are modified according to:
\begin{equation}\label{ourrule}
W^R_{kj}=W_{kj}+W_{kn}\frac{W_{nj}}{W^{out}_n}.
\end{equation}
This procedure corresponds to adding to the rate the process of $k$
going to $n$ with the proper rate and then instantaneously going to
$j$ with the proper probability. A graphical example of the
application of the method is shown in Fig. (\ref{figscheme}).

\subsection{Commutativity and adiabatic approximations}

A natural question at this point is whether the order of elimination
of the ``fast states'' matters for the resulting dynamics . In
presence of two fast ``linked'' states, say $n$ and $m$, such that
$\Delta t\ge \max\left\{(W_n^{out})^{-1}\right\}$, $\Delta t\ge
\max\left\{(W_m^{out})^{-1}\right\}$ and $W_{nm}>0$ or $W_{mn}>0$, the
result of decimation could be different, in principle, depending on
the order of elimination of the ``fast states'', i.e. before $n$ and
them $m$ or viceversa.

To show that our procedure commutes in general and clarify the
connection with adiabatic approximations, let us rearrange the vector
$(P_1(t),P_2(t)\ldots)$ into two vectors $(\psi_1(t),\psi_2(t))$ where
$\psi_1$ is a vector containing the probabilities of the slow states
and $\psi_2$ contains the probabilities of the fast states. We rewrite
the master equation as:
\begin{equation}
\frac{d}{dt}\!\left(\begin{array}{c}\psi_1\\
\psi_2\end{array}\right)\!=\!\left(\begin{array}{cc}\hat{B}_{11} & \hat{B}_{12}\\
\hat{B}_{21} & \hat{B}_{22}\end{array}\right) \left(\begin{array}{c}\psi_1\\
\psi_2\end{array}\right)\!=\!\left(\!\begin{array}{c}\hat{B}_{11}\psi_1 + \hat{B}_{12} \psi_2\\
\hat{B}_{21}\psi_1 + \hat{B}_{22} \psi_2\end{array}\!\right).
\end{equation}

The adiabatic approximation corresponds to set $d\psi_2 / dt=0$, that is
\begin{equation}\label{adcondition}
\hat{B}_{21}\psi_1+\hat{B}_{22}\psi_2=0.
\end{equation}

We can safely assume $\det \hat{B}_{22}\neq0$ since otherwise the
probabilities of the slow states would be zero at equilibrium. So we
can solve the above equation for $\psi_2$ and substitute it into the
equation for $\psi_1$:
\begin{equation}\label{eqadiabatic}
\frac{d}{dt} \psi_1 =(\hat{B}_{11} - \hat{B}_{12}\hat{B}^{-1}_{22}\hat{B}_{21})\psi_1 .
\end{equation}
Notice that Eq. (\ref{eqadiabatic}) preserves normalization since
\begin{equation}
\frac{d}{dt}\sum_{n}[ P_n(t)\in \psi_1(t)] =\frac{d}{dt} \sum_n P_n(t)=0
\end{equation}
in other words, within the adiabatic approximation there is no
probability flow between fast and slow states.  Eq.(\ref{eqadiabatic})
may thus be considered a {\it bona fide} reduced master equation for
the slow degrees of freedom, with the transition rates being
renormalized due to the effect of the fast states. Notice that, at
variance with fast variables elimination methods \cite{frankowicz},
the probabilities $\psi_1$ are not marginalized probabilities, meaning
that we didn't average over fast states. The probabilities of fast
states can be eventually reconstructed as a function of time after
solving the equations for the slow states.

To clarify the relationship between Eq. (\ref{eqadiabatic}) and our
method, notice that condition (\ref{adcondition}) is a linear set of
equations for the fast degrees of freedom. Let us consider the
solution obtained by the substitution method: we start eliminating a
particular fast state $j$ by computing its probability from the $j$-th
equation:
\begin{equation}
P_j=\frac{\sum_{i\neq j}W_{ij}P_i(t)}{W^{Tot}_j}
\end{equation}

It should be clear at this point that, by substituting the above
expression in the evolution equation of a fast state, the original
rule of Eq. (\ref{ourrule}) is retrieved; the same procedure can be
iterated to solve the probabilities of all the fast states and
substitute it into the remaining equations.  The conclusion is that
our method is equivalent to solve the condition (\ref{adcondition})
using the substitution method. But since the solution of that
condition is unique, the resulting master equation will be
Eq.(\ref{eqadiabatic}), independently on the order of elimination of
the fast variables.

\section{One dimensional examples}\label{onedsection}

Now we apply our decimation procedure to some one
dimensional random processes.  The first example is the well known
problem of the double well; in this case the decimation procedure can
be carried out analytically and predict the correct transition rate
between the two minima.  The second example is a one dimensional
random walk with defect.

\subsection{Double well potential}

Let us consider the problem of a random walk in a double-well
potential. This example is paradigmatic both in physics and in kinetic
chemistry, in the latter case the one-dimensional axis represents the
reaction coordinate and one is interested the rate of jumping from one
minimum to the another, i.e. of the reaction to occur. Again, we
assume that this axis is subdivided in discrete states $\{n\}$. We can
introduce the potential $V_n$, which determines the following
transition probabilities:
\begin{equation}\label{potential}
W_{j\rightarrow j\pm 1}= \exp\left[\frac{\beta}{2}(V_j-V_{j\pm 1}) \right]
\end{equation}
\begin{figure}[h]
\begin{center}
\includegraphics[width=8.3cm]{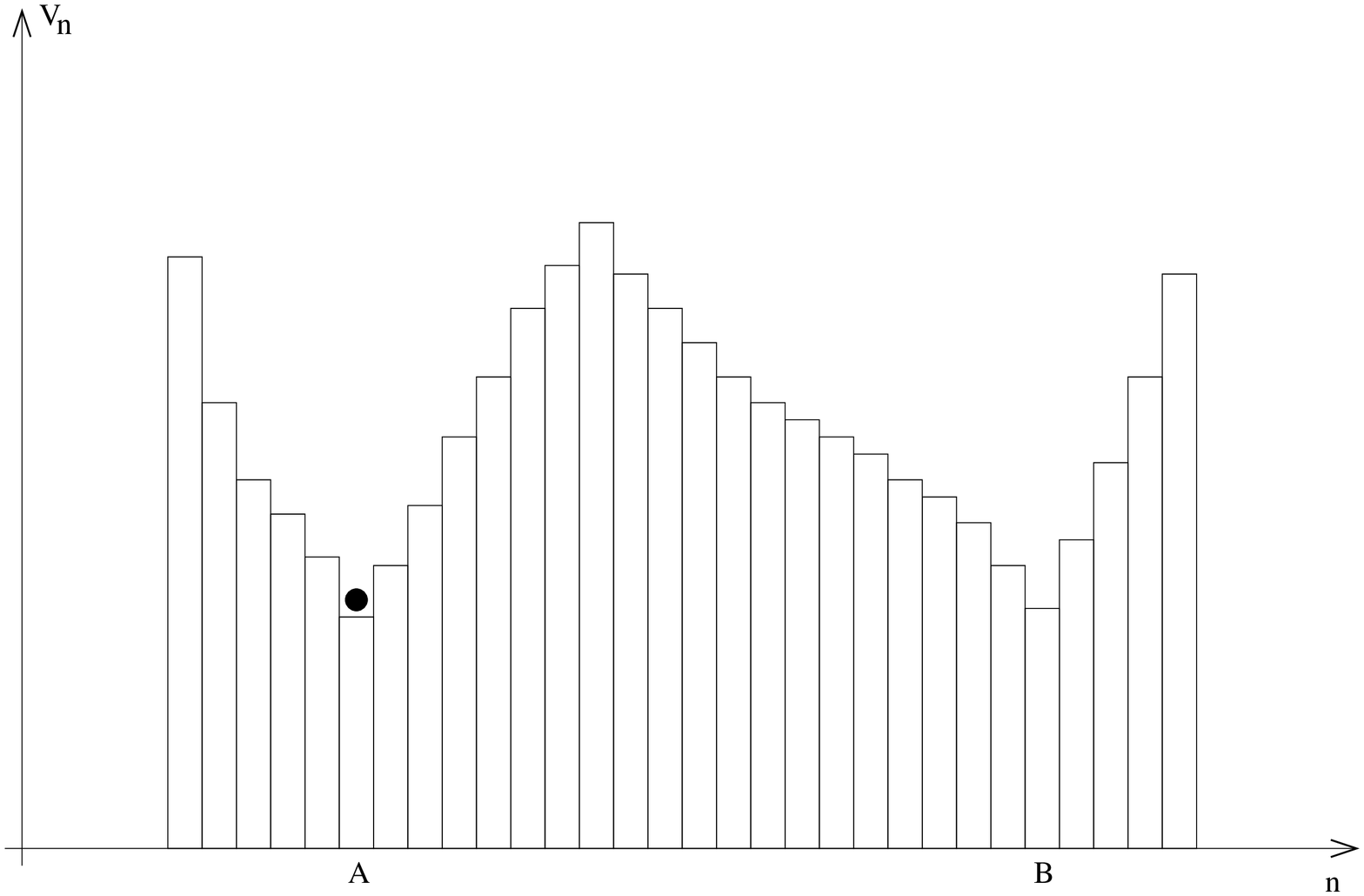}
\end{center}
\caption{Particle in a discrete double well potential. $A$ and $B$ are
the states corresponding to the two minima of the
potential.}\label{figuradw}
\end{figure}
being $\beta$ the usual Boltzmann factor.  Notice that the above
transition rates ensure that at equilibrium the probability of being
in state $n$ is proportional to $\exp(-\beta V_n)$. We do not restrict
ourself to a particular choice of the potential, we just assume that
the potential has two minima which we denote with $n=A$ and $n=B$, as
sketched in Fig.(\ref{figuradw}). As usual, we are free to choose a
value of $\Delta t$ and eliminate all the states whose inverse
out-rate is smaller than $\Delta t$:
\begin{equation}\label{surviving}
\Delta t > 1/W_j^{out}=
\left(e^{\frac{\beta}{2}(V_j-V_{j-1})}+ 
e^{\frac{\beta}{2}(V_j-V_{j+1})}\right)^{-1}
\end{equation}

By increasing $\Delta t$, the two minima are the last two surviving
states since in a minumum both the exponentials in
Eq.(\ref{surviving}) have negative arguments. This means that with a
proper choice of the time scale we can end up with a two state Markov
chain and calculate the transition rate between the two
minima. 

By alling $N$ the number of states between $A$ and $B$, the
transition rate writes:
\begin{equation}\label{renortrans}
W^R_{A\rightarrow B}=\frac{W_{A\rightarrow A+1}  W_{A+1\rightarrow
    A+2}\ldots 
 W_{A+N\rightarrow B}}{W^{out,R}_{A+1}W^{out,R}_{A+2}\ldots W^{out,R}_{A+N}}.
\end{equation}
In the denominator we indicate with the notation $W^{out,R}_i$ the
total out-rate of state $i$ to remember that this out-rate changes
when a neighboring state is eliminated. This means that the product
of out-rates has to be evaluated by eliminating the states one after the
other.  To evaluate the expression (\ref{renortrans}), we make use of
the following equality, which is demonstrated in the Appendix:
\begin{equation}\label{equality}
\prod_{h=i+1}^{i+N}W_h^{R}=
\sum_{j=0}^{N}\left(\prod_{k=i+1}^{i+j} W_{k\rightarrow
  k-1}
\prod_{k=i+j+1}^{i+N} W_{k\rightarrow k+1}\right)
\end{equation}
with the convention that products of less than one terms are always
equal to one, $\prod_{k=1}^0 W_k=1$. Substituting
(\ref{potential}) and (\ref{equality}) into (\ref{renortrans}) one
obtains:
\begin{eqnarray}
W^R_{A\rightarrow B}=
\frac{\exp[\frac{\beta}{2}(V_A-V_B)]}
{\sum_{j=0}^N  \exp[\frac{\beta}{2}(\!V_{A+j}\!-\!V_{A}\!+\!V_{A+j+1}\!-\!V_{B})]}=
\\
\!\!\!\!\!\!=\frac{1}{\sum\limits_{j=0}^N \exp[\frac{\beta}{2}(V_{A+j}\!+\!V_{A+j+1})]}
\sim \frac{1}{\sum\limits_{j=0}^N \exp(\beta V_{A+j})}\nonumber
\end{eqnarray}
where in the last step we assumed that the potential does not change
much between adjacent states. The last expression is basically
the well known result for the transition rate in a double well
potential in a continuous system described by a Fokker-Planck
\cite{doublewellref}.

\subsection{Random walk with defects}

Considered now a random walker in a one dimensional geometry with
periodic boundary conditions; the number of its possible states is
fixed to $N=10^3$. The transition rate between two adjacent states is
$W_{n,n-1}=W_{n,n+1}=1$ except for a fraction $p=0.2$ of the states
(the ``defects''), having $W_{n,n-1}=W_{n,n+1}=20$. The defects are
placed at random at the beginning and kept the same in all the
simulations (quenched disorder).  In solid state physics the master
equation of this system corresponds to the Schr\"odinger equation (at
imaginary time and with disorder) in the limit of tight binding
approximation \cite{AM}.

Clearly one has $W^{out}=2$ for the normal
states and $W^{out}=40$ for the defects. This means that the defects
are the fast states and we will test what happens when eliminating
them from the description. For comparison, we will see also the
differences with the ideal case without defects, that is
$W_{n,n-1}=W_{n,n+1}=1$ for all the states.

It is easy to analytically calculate the eigenvalues in the case
without defects \cite{AM}:
\begin{equation}
\lambda_j= 2\left[\cos\left(\frac{j}{N}\right)-1\right]\qquad j=0,\pm
1,\pm 2\ldots\pm\left(\frac{N}{2}-1\right).
\end{equation}

Notice that all the eigenfunction are non-localized, which in solid
state physics corresponds to a conductive system. On the contrary, in
the case with defects (for any positive value of $p$) the
eigenfunctions are localized, i.e. one has a transition from a metal
to an insulator \cite{LM}.

\begin{figure}[h]
\begin{center}
\includegraphics[width=8.3cm]{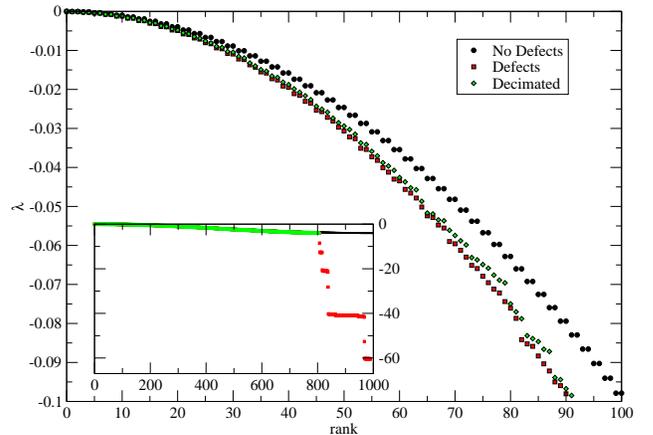}
\end{center}
\caption{Eigenvalues, listed in decreasing order of a random walk with
  defect. The number of state is $N=10^3$. Black circles: no defects,
  the jump rate is $W_{i,i+1}=W_{i,i=1}=1$. Red squares: with
  probability $p=.2$ the sites have defects and their jump probability
  is $W^d_{i,i+1}=W^d_{i,i=1}=20$. Green diamonds: random walk with
  defects after applying the decimation scheme to all the fast
  states. All the eigenvalues are real. The difference between the
  figure and the inset is just the axis scale.  Notice in the main
  figure that the first $100$ eigenvalues of the decimated problem
  follow very closely the case with the defects. In the inset the
  eigenvalues corresponding to the fast states are evident; notice
  that their number is the same as the average number of defects
  $p*N$.}\label{figeigenvalues}
\end{figure}

In Fig.(\ref{figeigenvalues}) we show the eigenvalues of the master
equation in the three cases: random walk without defects, with defects
and after decimation. Being the master equation linear, the
eigenvalues spectrum contain all the informations about the dynamics.
The spectrum shows an interesting property: the most negative
eigenvalues are much larger in modulus and correspond to fast decaying
eigenfunctions concentrated on the defects. It is clear from the
figure that the effect of our algorithm is to eliminate these
eigenvalues (and the corresponding eigenfunctions).

\begin{figure}[h]
\begin{center}
\includegraphics[width=8cm]{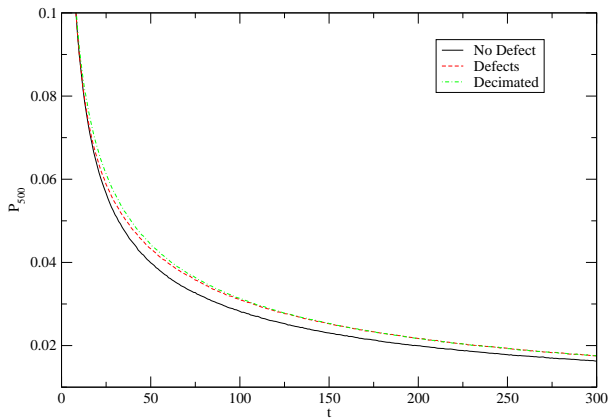}
\end{center}
\caption{Correlation functions. The system is the same of
Fig.(\ref{figeigenvalues}) prepared in the initial state $i=500$ and
the probability of being in state $i$, averaged over $10^7$
realization, is plotted as a function of time. The three curves are:
(black, continuous) the system without defects, (red, dashed) system
with defects, and (green, dot-dashed) the decimated
system.}\label{figcorr}
\end{figure}

It is clear in the figure that the 100 eigenvalues smaller in modulus,
corresponding to the slower dynamics of the system, are very similar
in the model with defects and the decimated one. A further check comes
from the correlation functions. We performed the simulations starting
from a slow state $i=500$ and plotted in Fig. \ref{figcorr} the
probability of being in the state $i$ as a function of time in the
three cases we considered. After an initial time, the correlation
functions for the decimated model and the model with defects are very
close to each other.

\section{Enzymatic reactions}\label{enzymesection}

In this section we apply our method to a model of enzymatic
reactions. Enzymatic reactions are widely studied in kinetic chemistry
and many methods have been derived to predict the steady-state
kinetics in cases in which the reaction is close to an equilibrium, or
at least in a non equilibrium steady state \cite{segel}.  To exemplify
the relevance of our method in this case, we will consider in the
following the simplest model of an enzymatic reaction:

\begin{equation}\label{reaction}
E + S \stackrel{\xrightarrow{k_1}}{\xleftarrow[k_{-1}]{}} ES
\xrightarrow{k_2} E + P
\end{equation}
where as usual $E$ denotes the enzyme, $S$ the substrate and $P$ the
reaction product. This correspond to the following master equation:
\begin{eqnarray}\label{meenzyme}
\frac{d}{dt}P(N_S,N_E)=k_1(N_S-1)(N_E-1)P(N_S\!-\!1,N_E\!-\!1)
\nonumber\\\!\!\!+k_{-1}(N_E^T-N_E)+P(N_S\!+\!1,N_E\!+\!1)+\quad\\
+k_2(\!N_E^T\!-\!N_E\!)P(N_S,N_E\!+\!1)-P(N_S,N_E)W^{out}(N_E,N_S)\nonumber
\end{eqnarray}
where we called $N_S$ the number of free (non binded) substrate
molecules, $N_E$ the number of free enzymes molecules, and $N_E^{T}$
the total number of enzyme molecules. We also introduced, as usual,
the total out-rate of a general state:
\begin{eqnarray}\label{outratesenzyme}
W^{out}(N_E,N_S)=k_1 N_E N_S +(k_{-1}+k_2)
(N_E^T-N_E)=\nonumber\\(k_{-1}+k_2)N_E^T+N_E[k_1 N_S-(k_1+k_2)].\quad
\end{eqnarray}

The quantity of interest is usually the production rate, that is the
velocity of the rightmost reaction as a function of the concentration
of enzymes and substrate.  It is easy to estimate it in the
quasi-equilibrium approximation, that is when $k_{-1}\gg k_2$, meaning
that the leftmost part of the reaction can be considered at
equilibrium. Another thing one can assume is steady state dynamics:
the calculation is simple even if the complex is not at equilibrium
but the concentration of the complex $ES$ does not vary much with
time, $d[ES]/dt\approx 0$. The steady state approximation is known to
hold very well when the concentration of enzymes is much smaller than
that of substrate.

In this case the reaction velocity follows the Michaelis-Menten formula:
\begin{equation}\label{mmformula}
v=\frac{k_2 E_T S}{K_m+S}
\end{equation}
where $K_m=(k_{-1}+k_2)/k_1$. In order to apply our method, we start
from Eq. (\ref{outratesenzyme}).  The expression for the out-rates
suggests two limiting cases in which there is separation of scales
between fast and slow states. The first case is $k_1 N_S \ll
(k_{-1}+k_2)$. In this case, states having $N_E>0$ have a much greater
out-rate than those with $N_E=0$ and can be coarse-grained. Physically
it corresponds to the situation in which the complex is efficiently
processed, so that its concentration is always very close to zero. In
this case the application of our method yields:
\begin{equation}\label{lowdensityformula}
W_{N_S^T\rightarrow N_S^T-1}=\frac{k_2 N_E^T N_S^T}{K_m}
\end{equation}
Notice that we wrote the rate in term of $N_S^T$ (the total number of
substrate molecules) and not $N_S$ since with the coarse graining we
grouped together states with bounded and free substrates as long as
the total number is the same.  Notice also that this rate correctly
corresponds to the Michaelis Menten formula, eq. (\ref{mmformula})
taken in the limit $N_S/K_m\ll 1$. The new information provided by our
method is that in this limit the linear dependence on the number of
substrates is valid also beyond the steady state approximation, that
is when the enzymes concentration becomes large.

The second limiting case is $k_1 N_S \gg (k_{-1}+k_2)$. In this case,
all states having $N_E>0$ and $N_S>0$ are fast. This means that, for a
fixed $N_S^T$, the slow state is characterized by
$N_E=\max(0,N_E^T-N_S^T)$. When $N_E^T<N_S^T$ we retrieve again the
steady state result:
\begin{equation}\label{highdensityformula}
W_{N_S^T\rightarrow N_S^T-1}=k_2 N_E^T
\end{equation}
The above expression, again, is consistent with the Michaelis-Menteen
equation, this time in the limit $N_S/K_m\ll 1$ when the reaction rate
does not depend anymore on the substrate concentration. On the other
hand, when $N_E^T>N_S^T$ one has:
\begin{equation}\label{highdensitynostat}
W_{N_S^T\rightarrow N_S^T-1}=k_2 N_S^T.
\end{equation}
This last case is radically different with the steady state
prediction: we have a linear dependence on the number of substrate
molecules while Michaelis-Menteen formula predicts a rate which is
independent on $N_S^T$. The physical reason is that this is a case in
which the complex never reaches a steady state: on a fast timescale,
order $k_1^{-1}$ the free substrate is completely converted into the
complex $ES$. Then it is depleted on much slower times, and the
depletion speed is limited by the lower between the enzyme and the
substrate concentration.

\begin{figure}[h]
\begin{center}
\includegraphics[width=8cm]{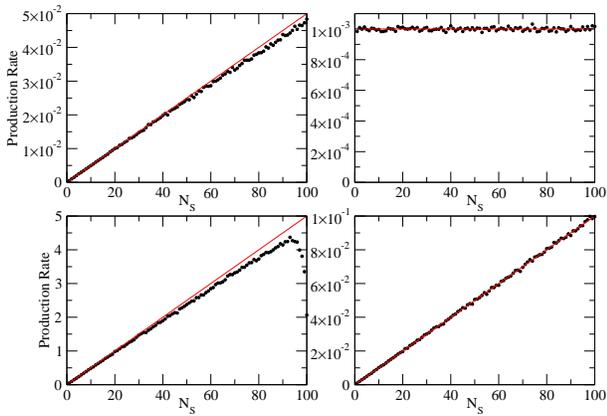}
\end{center}
\caption{Simulations of the master equation
corresponding to the enzymatic reaction of Eq.(\ref{reaction}). In all
simulations we start with $N_S=100$ molecules of free substrate and
let the system evolve: the rate as a function of the number of
molecules is evaluated by averaging over $10^4$ realizations of the
process. The number of enzymes is $N_E^T=1$ in the top figures and
$N_E^T=100$ in the bottom figures. The reaction rates are (left)
$k_1=10^{-3}$, $k_{-1}=k_2=1$ and (right) $k_1=1$,
$k_{-1}=k_2=10^{-3}$. The lines in the black and white versions
of the right figures are barely visible since the points fall very
close to them.}\label{figenzim}
\end{figure}

In Fig. (\ref{figenzim}) we plot simulations of the reaction
(\ref{reaction}) in four different situations. The top left figure
corresponds to the situations in which $N_E^T\ll N_S^T$ and $k_1 N_S
\ll (k_{-1}+k_2)$ (numerical values of the parameters are in the
figure caption), together with the prediction of
Eq. (\ref{lowdensityformula}), while on the top-right we show
$N_S^T\ll N_E^T$ and $k_1 N_S \gg (k_{-1}+k_2)$ with the prediction of
Eq. (\ref{highdensityformula}). These two are the cases in which
steady state approaches are known to work and our approach simply
corresponds to the limiting cases of Michaelis-Menten formula. On the
bottom figures we show the same cases, but with $N_E^T\le
N_S^T$. Notice that the left one correspond to the top/left one,
rescaled with the higher number of enzymes; in particular, it is still
described by Eq. (\ref{lowdensityformula}). On the other hand, the
bottom-right is even qualitatively different from the corresponding
top figure. Eq.(\ref{highdensitynostat}) correctly describes the
behavior, while Michaelis Menten formula would predict the rate to be
independent of $N_S^T$.

\begin{figure}[t]
\begin{center}
\includegraphics[width=8cm]{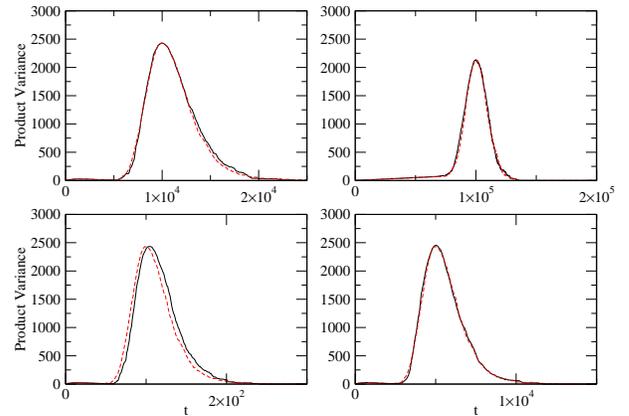}
\end{center}
\caption{Product variance as a function of time, each
averaged over $10^3$ realizations. The four figures correspond to the
same parameter choices of Fig. (\ref{figenzim}). Black continuous
lines correspond to simulation of the full master equation, red dashed
lines are simulations of the reduced processes defined by
Eq.(\ref{lowdensityformula}), (\ref{highdensityformula}) and
(\ref{highdensitynostat}).}\label{figvar}
\end{figure}

In order to study also the fluctuations of the process, we simulated
both the full master equation and the simplified processes defined by
Eq. (\ref{lowdensityformula}), (\ref{highdensityformula}) and
(\ref{highdensitynostat}). Then we compare in Fig. (\ref{figvar}) the
variance of the number of product molecules as a function of time. The
result of the simulations is that the fluctuations are almost
indistinguishable in the reduced processes: the intuitive reason is
that most of the fluctuations are related to the slow states and thus
unaffected by the coarse graining procedure.

We conclude this section with some remarks. The condition $k_1 N_S \ll
(k_{-1}+k_2)$ has a physical interpretation: it corresponds to the
situation in which the concentrations are very low. Indeed $k_1$ is
the only rate that depends on the volume since it is essentially
determined by the search time. Conversely, the condition $k_1 N_S \gg
(k_{-1}+k_2)$ corresponds to large concentrations. This means that all
the four cases we discuss are, in general, experimentally accessible.
This kind of approach may be useful in the study of enzymatic
reactions inside the cells. Such reactions are often characterized by
low number of molecules \cite{elowitz}, and probably the
Michaelis-Menten picture is appropriate even with a low number of
substrate molecules. On the other hand, one can study situations in
which concentrations are very high, or cases in which the changes that
can occur in the cell environment, for example as a response to a
rapidly varying external signal \cite{tiana}, can bring the reaction
outside the steady state regime and make a description of this kind
more appropriate.

\section{Conclusions}\label{conclsection}

In this paper we introduced a general method for the decimation of
``fast modes'' in systems evolving according to a Master equation via
a coarse graining procedure.  Our method is general, and somehow, in
the spirit of the renormalization group (RG) approach. Indeed, similar
approaches have been proposed in statistical mechanics for the study
of disordered systems \cite{igloi}. At variance with the RG, we do not
aim at reaching a fixed point by repeating the decimation
procedure. In general, the method is applied only once: a sensible
value of the parameter $\Delta t$, which selects the coarse graining
level, may be easily chosen by looking at the magnitude of the total
out rates $W_n^{out}$.

We show that the procedure is commutative and brings to consistent
results for a general master equation, independently of the
dimensionality. However, for the sake of simplicity, we discussed in
details low dimensional examples, where the method brings also to
analytical predictions.  In the discrete version of the double well
potential, the decimation procedure is able to reproduce the well
known result for the transition time. A numerical analysis shows that
the method gives a very good approximation of the original system in
the case of a random walk with defects. In the case of an enzymatic
reaction, we show that the method allows for predictions only in some
well defined limiting cases. However, in these cases, the prediction
are more general than those obtained within a steady-state
approximation.

Let us conclude with a short comparison with other approaches to
multiscale master equations. Our method is more similar in spirit to
projection methods\cite{munsky,peles} than to other quasi-steady
states methods \cite{haseltine,cao}. The reason is that the result of
our procedure is still a master equation while the other method
describe the fast dynamics in a different way (with a differential or
Langevin equation).  The main difference between our method and the
finite state projection is that our procedure is ``local'' (we
consider fast and slow states) while projection methods generally
consider eigenvalues and eigenvector of the transition
matrix. Projection methods are usually of more general applicability,
however our procedure may allow a more transparent interpretation of
the surviving states, as we show in the examples we considered.

On the other hand, fast variables elimination methods
\cite{frankowicz,pineda} aim at writing an evolution equation for the
probability of slow states summed over the probability of the fast
ones. In this way one loses informations about the fast states and, if
there is separation of scales, may write a closed equation for the
slow ones. We show that our approach does not imply such coarse
graining. In fact, no information about the fast states is lost in our
case and their dynamics can be reconstructed after solving the problem
involving the slow states only.

There are also analogies between our method and some coarse graining
procedure that have been proposed in the field of complex networks. In
this field, a relevant problem is the identification of communities
\cite{newman} and a possible way to do it is to consider a diffusion
process on the network and try to coarse grain the graph while keeping
the long timescale properties of this dynamics
\cite{reichardt,gfeller}. The difference is that in these cases the
procedure allows for a spatial simplification of the links, while our
decimation procedure in a system whose states are seen as elements of
a graph brings a suppression of the fast states but without a relevant
simplification of the surviving connections.

\section*{Appendix: proof of equality (\ref{equality}) and commutativity in $1D$}

In this appendix we prove that, when decimating a cluster of $N$
consecutive states, the product of their out rates can be written as:
\begin{equation}\label{toprove}
\prod_{h=i+1}^{i+N}W_h^{out,R}=\sum_{j=0}^{N}\left(\prod_{k=i+1}^{i+j}
  W_{k\rightarrow k-1} \prod_{k=i+j+1}^{i+N} W_{k\rightarrow
  k+1}\right)
\end{equation}
remembering the convention that the product of less than one object is
equal to one, and it is independent of the order of decimation. 

\begin{figure}[h]
\begin{center}
\includegraphics[width=8.3cm]{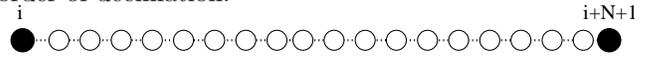}
\end{center}
\caption{Two states ($i$ and $i+N+1$, black) separates by a cluster of
N states (white) to be decimated.}\label{figapp}
\end{figure}

We will show that the above formula holds when decimating states in
consecutive order, say from state $i+1$ one after the other to state
$i+N$ (see Fig.(\ref{figapp})), remembering that the result does not
depend on the order of decimation due to the commutative property.
We will demonstrated it by induction: first of all, the above formula is
obviously true for $N=1$:

\begin{eqnarray}
&&W^{out,R}_{i+1}=\sum_{j=0}^{1}\left[\left(\prod_{k=i+1}^{i+j}W_{k\rightarrow
k-1}\right)\left(\prod_{k=i+j+1}^{i+1}W_{k\rightarrow
k+1}\right)\right]=\nonumber\\&&W_{i+1\rightarrow i+2}+W_{i+1\rightarrow i}
\end{eqnarray}
Now we show that if the equality holds for a given value of $N$, then
it holds also for $N+1$:
\begin{widetext}
\begin{eqnarray}
&&\prod_{h=i+1}^{i+N+1}W_H^{out,R}=\sum_{j=0}^{N}\left(\prod_{k=i+1}^{i+j}
  W_{k\rightarrow k+i}\prod_{k=i+j+1}^{i+N} W_{k\rightarrow
  k+1}\right)W^{out,R}_{i+N+1}=\nonumber\\
  &&=\sum_{j=0}^{N}\left(\prod_{k=i+1}^{i+j} W_{k\rightarrow
  k+i}\prod_{k=i+j+1}^{i+N} W_{k\rightarrow
  k+1}\right)\left(W_{i+N+1,i+N+2}+\frac{\prod_{k=i+1}^{i+N+1}W_{k\rightarrow
  k-1}}{\prod_{k=i+1}^{i+N}W_k^{out,R}}\right)=\nonumber\\&&
  =\sum_{j=0}^{N}\left(\prod_{k=i+1}^{i+j} W_{k\rightarrow
  k+i}\!\prod_{k=i+j+1}^{i+N+1} W_{k\rightarrow
  k+1}\right)\!+\!\prod_{k=i+1}^{i+N+1}W_{k\rightarrow k-1}
  =\sum_{j=0}^{N+1}\left(\prod_{k=i+1}^{i+j} W_{k\rightarrow
  k+i}\prod_{k=i+j+1}^{i+N+1} W_{k\rightarrow k+1}\right)\nonumber
\end{eqnarray}
\end{widetext}
which is exactly the same expression of eq.(\ref{toprove}) for
$N\rightarrow N+1$. This completes our proof.

\begin{acknowledgments}
We are grateful to M. Cencini and A. Puglisi for useful remarks and a
detailed reading of the manuscript.  A.V. wishes to thank {\it
Universitad de las Islas Baleares} (Palma de Mallorca, Spain) for
hospitality during the first stage of this work. S.P. wishes to thank
A.D. Jackson for stimulating discussions and help with the commutativity
argument.
\end{acknowledgments}

\end{document}